\documentclass[twocolumn,british,prl,aps,showpacs,superscriptaddress]{revtex4-1}
\usepackage[letterpaper]{geometry}
\usepackage{url}
\usepackage{graphicx}
\usepackage{amsmath,bbm}
\usepackage{amssymb}
\usepackage{bbold,soul}
\usepackage{esint}
\usepackage{epstopdf}
\usepackage{subfigure}
\usepackage{color}
\usepackage{babel}
\usepackage{graphics}
\usepackage{epsfig}


\newcommand{\incoh}{\text{rnd}}

\newcommand{\KB}{}

\begin{document}

\title{Visibility-based hypothesis testing using higher-order optical interference}
\author{Micha\l{} Jachura}
\email{michal.jachura@fuw.edu.pl}
\affiliation{Faculty of Physics, University of Warsaw, ul. Pasteura 5, 02-093 Warszawa}
\author{Marcin Jarzyna}
\affiliation{Centre of New Technologies, University of Warsaw, ul. Banacha 2c, 02-097 Warszawa}
\author{Micha\l{} Lipka}
\affiliation{Faculty of Physics, University of Warsaw, ul. Pasteura 5, 02-093 Warszawa}
\author{Wojciech Wasilewski}
\affiliation{Faculty of Physics, University of Warsaw, ul. Pasteura 5, 02-093 Warszawa}
\author{Konrad Banaszek}
\affiliation{Faculty of Physics, University of Warsaw, ul. Pasteura 5, 02-093 Warszawa}
\affiliation{Centre of New Technologies, University of Warsaw, ul. Banacha 2c, 02-097 Warszawa}
\date{\today}


\begin{abstract}
Many quantum information protocols rely on optical interference to compare datasets with efficiency or security unattainable by classical means. Standard implementations exploit first-order coherence between signals whose preparation requires a shared phase reference. Here, we analyze and experimentally demonstrate binary discrimination of visibility hypotheses based on higher-order interference for optical signals with a random relative phase. This provides a robust protocol implementation primitive when a phase lock is unavailable or impractical. With the primitive cost quantified by the total detected optical energy, optimal operation is typically reached in the few-photon regime.
\end{abstract}

\maketitle

\begin{figure}[b]
\begin{center}
\includegraphics[width=0.8\columnwidth]{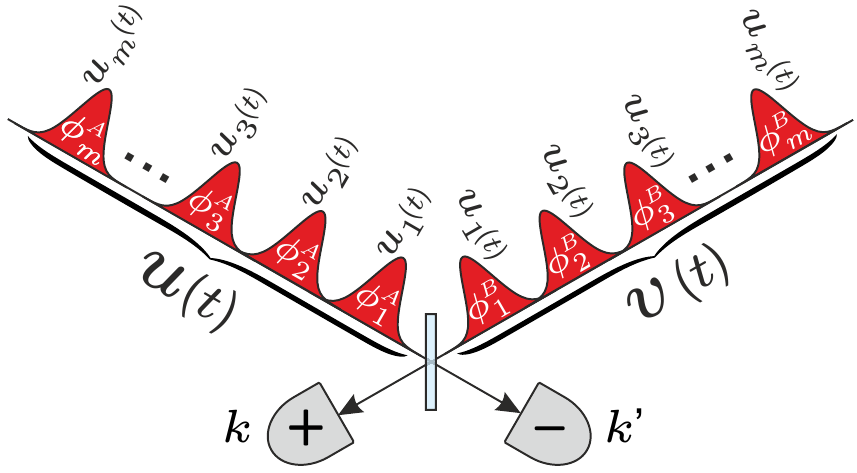}
\caption{An interferometric primitive for visibility-based hypothesis testing with phase-keyed signals. Input data in possession of two parties $A$ and $B$ are mapped onto phase patterns $\phi_1^A, \phi_2^A, \ldots, \phi_m^A$ and $\phi_1^B, \phi_2^B, \ldots, \phi_m^B$ used to modulate sequences of $m$ pulses described by a family of normalized temporal waveforms $u_1(t), u_2(t), \ldots, u_m(t)$. Generated optical signals can be viewed as prepared in collective modes described respectively by
$u(t) = \sum_{j=1}^{m} u_j(t) \exp(i\phi_j^A) /\sqrt{m}$ and $v(t) = \sum_{j=1}^{m} u_j(t) \exp(i\phi_j^B) /\sqrt{m}$. The signals are brought to interference at a $50/50$ beam splitter whose output ports are monitored by photodetectors. The outcome of a single repetition of an interferometric measurement is a pair of integers $k,k'$ specifying the number of counts registered by each of the detectors over the duration of the signals.}
\label{Fig:scheme1}
\end{center}
\end{figure}

Optical systems, in addition to being the workhorse of modern telecommunication, provide a natural platform to implement quantum-enhanced protocols for information transfer and processing between distant parties. Quantum strategies can provide authentication or reduce the communication complexity of certain tasks, in which large distributed datasets need to be processed to infer a relatively small amount of information \cite{Brassard2003, Buhrman2010}. Examples include  quantum digital signatures \cite{Gottesman2001} and quantum fingerprinting \cite{Buhrman2001}. These protocols share a primitive  which consists in imprinting the input data onto the modal structure of transmitted fields, e.g.\ in the form of phase patterns, and interfering the received signals, as shown in Fig.~\ref{Fig:scheme1}. Different hypotheses, e.g.\ the instances of identical and unequal inputs, are mapped onto distinct ranges of the interference visibility, which can therefore serve as the basis for hypothesis testing. Strikingly, optical signals sufficient to realize the quantum scheme may not have the capacity to carry information necessary to implement the classical protocols with the matching confidence level. This enhancement, stemming from the interplay between wave and particle properties of light exploited in quantum protocols, can advantageously change the scaling of resources required to perform the task \KB{as well as ensure security}.

As recently pointed out \cite{Arrazola2014, Arrazola2014a} and demonstrated experimentally \cite{Clarke2012, Collins2014, Xu2015, Guan2016}, the protocol primitive described above can be realized efficiently with coherent light beams and first-order interference. This implementation uses laser light sources and is robust against attenuation introduced by optical channels  transmitting the signals, but it requires phase stability between the sending parties. In certain scenarios a shared phase reference may be unavailable or very difficult to furnish. An alternative may be to resort to Hong-Ou-Mandel interference between single photons which has been exploited in proof-of-principle demonstrations of quantum communication complexity protocols \cite{Horn2005, Ekert2006}. However, a practical implementation may require single photon sources with long coherence times and would be inefficient for high channel attenuation. The latter impairment affects also a realization based on weak classical states with a random global phase \cite{Jachura2017}.

\begin{figure*}[htbp!]
\begin{center}
\includegraphics[scale=0.6]{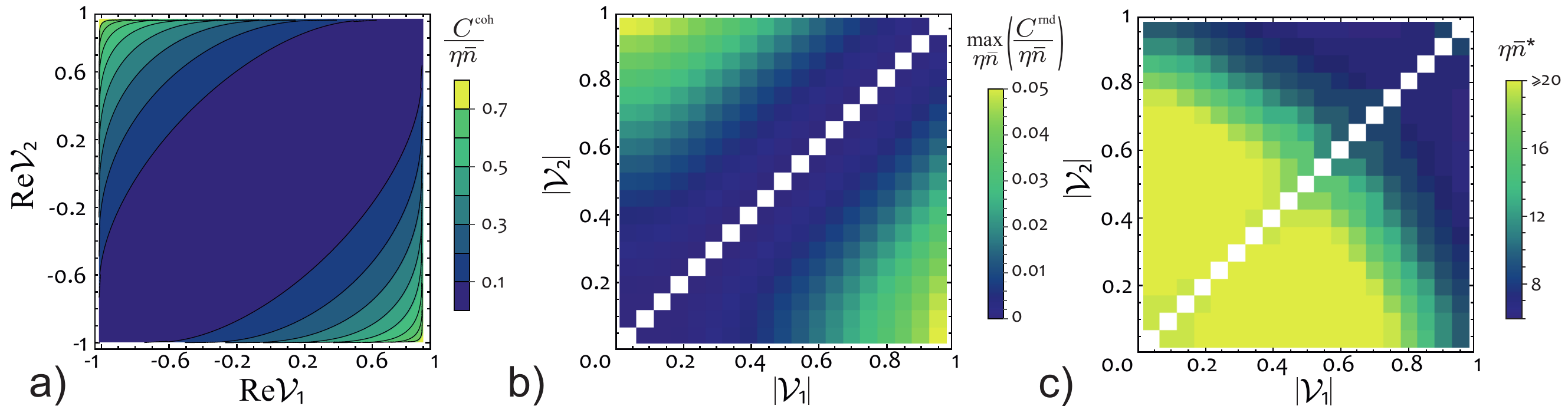}
\caption{\KB{Discrimination between a pair of hypotheses encoded in interference visibilities ${\cal V}_{1},{\cal V}_{2}$ for the coherent and the random-phase scenario. (a) Information gained from a single photodetection event for a fixed phase between interfering signals. (b) Maximum information per one detected photon $C^{\incoh}/(\eta\bar{n})$ for signals with a random global phase. (c) Optimal $\eta\bar{n}^\ast$ maximizing the ratio $C^{\incoh}/(\eta\bar{n})$. White squares on the diagonal in (b), (c) represent the case $|{\cal V}_1|=|{\cal V}_2|$ when the two hypotheses are indistinguishable.}} \label{chernoffymapa}
\end{center}
\end{figure*}

In this paper we present a strategy to carry out optical hypothesis testing based on the visibility of higher-order interference between classical fields with a random relative phase. This approach concurrently benefits from conventional optical signal generation techniques, removes the need for a shared phase reference, and ensures robustness against channel attenuation. The  performance is characterized using average error probability, whose asymptotic behavior is investigated with the help of a refined Chernoff bound \cite{Cover1991, Bahadur1960}. Interestingly, we show that when the protocol cost is quantified in terms of the total transmitted optical energy, the optimal strategy is to realize multiple repetitions of the interference visibility measurement in the few-photon regime with a determination of the complete photocount statistics.



Let us first consider interference between two mutually coherent optical signals. Each signal has the form of a pulse sequence depicted in Fig.~\ref{Fig:scheme1} and carries optical energy $\bar{n}/2$ expressed in photon number units.
The receiver combines the signals on a balanced beam splitter. The time-integrated light intensity at the two output ports of the beam splitter labeled with indices `$+$' and `$-$' can be written as $I^{\pm}({\cal V})  = {\eta \bar{n}} (1 \pm \text{Re} {\cal V})/2$, where $\eta$ is the channel transmission for each of the signals \cite{suplement}. Here ${\cal V}$ is the interference visibility which carries information about the relation between the input datasets. In the ideal case it is equal to the overlap ${\cal V} = \int dt \, u(t) v^\ast(t)$ between the normalized waveforms $u(t)$ and $v(t)$ describing the two received signals. For identical inputs 
${\cal V}=1$, which corresponds to completely destructive interference at the `$-$' output port. Hence registering a photocount at that port unambiguously indicates that the inputs were unequal. This observation underpins the quantum fingerprinting protocol which aims at deciding whether datasets in posession of two parties are identical or different while revealing the smallest possible amount of information to the external referee. The protocol employs classical error correction to guarantee that for any pair of unequal inputs the visibility remains below a certain threshold value. Given that experimental imperfections, such as detector dark counts and misalignment of optical beams, lower the effective visibility \cite{suplement}, the hypotheses of identical or unequal inputs correspond to two distinct ranges of the visibility parameter separated by a gap. In order to perform a practical test between these two hypotheses one needs to devise a decision rule based on the measured photocount statistics.

Let the detectors at the the output ports of the the beam splitter be able to resolve up to $K$ photocounts over the signal duration. The probability $p_k^\pm$ of registering $k$ photocounts on one detector reads
$p_{k}^{\pm}({\cal V}) = \exp[-I^\pm({\cal V})] {[I^{\pm}({\cal V})]^{k}}/{k!}$
for $k = 0,1,\ldots, K-1$ and $p_{K}^{\pm}({\cal V}) = 1 - \sum_{k=0}^{K-1} p_{k}^{\pm}({\cal V})$. Non-unit efficiency of the detectors can be included in the channel transmission $\eta$. Suppose now that the signal pairs are received with a promise that the visibility takes only one of two equiprobable values ${\cal V}_1$ or ${\cal V}_2$. For the fingerprinting protocol one value corresponds to identical inputs, while the second one can be taken as the highest visibility occurring in the case of unequal inputs.
The task is to discriminate between the two visibility hypotheses on the basis of the photocount sample collected in $N$ repetitions of the interferometric measurement. The probability $\varepsilon$ of erroneously identifying the actual visibility is upper bounded by the so-called Chernoff bound $\varepsilon \le \exp(-NC)/2$ \cite{Cover1991}, where $C$ stands for the Chernoff information given explicitly by
\begin{equation}
C=-\log\left[\min_{0 \le \alpha \le 1}\left(\sum_{k,k' = 0}^{K}
[P_{kk'}({\cal V}_1)]^{\alpha}[P_{kk'}({\cal V}_2)]^{1-\alpha}\right)\right].
\label{Eq:ChernoffExponentDef}
\end{equation}
In the above expression, summation is carried out over all possible measurement outcomes, which in our setup have the form of two integers $k$ and $k'$ specifying the number of counts registered by individual detectors, and $P_{kk'}({\cal V})$ \KB{denotes} the probability of obtaining a specific combination $kk'$ for the visibility ${\cal V}$.

For the coherent signal scenario considered so far, the probability of registering respectively $k$ and $k'$ counts has the product form $P^{\text{coh}}_{kk'}({\cal V}) = p^{+}_{k}({\cal V}) p^{-}_{k'}({\cal V})$. Assuming full photon number resolution with $K\rightarrow \infty$, the Chernoff information can simplified to
\begin{multline}
C^{\text{coh}} = \eta\bar{n} \left( 1 - \frac{1}{2} \min_{0 \le \alpha \le 1} [(1+\text{Re} {\cal V}_1)^\alpha (1+\text{Re} {\cal V}_2)^{1-\alpha} \right. \\
\left.\vphantom{\frac{1}{2}} + (1-\text{Re} {\cal V}_1)^\alpha (1-\text{Re} {\cal V}_2)^{1-\alpha} ] \right)
\KB{.}
\end{multline}
It is seen that the Chernoff \KB{information} is proportional to the received optical energy $\eta\bar{n}$. The proportionality factor given by the ratio $C^{\text{coh}}/(\eta\bar{n})$ can be interpreted as the amount of information gained from the detection of one photon. In Fig.~\ref{chernoffymapa}(a) we depict this factor as a function of the \KB{real parts of} visibilities $\text{Re} {\cal V}_1$ and $\text{Re} {\cal V}_2$. Generally, it pays off to maintain a large distance between the visibilities with the maximum information attained for ${\cal V}_1 = - {\cal V}_2= \pm 1$.

The above picture becomes much more nuanced if the sending parties have no access to a shared phase reference, which implies that the signals arrive with a random relative phase. However, in each individual realization the signals are described by coherent waveforms whose overlap is given by ${\cal V}$ up to \KB{an} overall phase factor. In such a scenario, the joint photocount distribution reads
\begin{equation}
P^{\incoh}_{kk'}({\cal V}) =  \int_{0}^{2\pi}  \frac{d\varphi}{2\pi} \, p^{+}_{k}(e^{i\varphi}{\cal V}) p^{-}_{k'}(e^{i\varphi}{\cal V}).
\label{Eq:probability}
\end{equation}
The explicit analytical expression for $P^{\incoh}_{kk'}({\cal V})$ \KB{is derived in Supplemental Material \cite{suplement}}. Obviously, after averaging over the global phase only the absolute value $|{\cal V}|$ of the visibility parameter is relevant. The above probability distribution can be used to calculate the respective Chernoff information $C^\incoh$ according to Eq.~(\ref{Eq:ChernoffExponentDef}). As before, the ratio $C^\incoh/(\eta\bar{n})$ has the interpretation of the amount of information gained per one received photon.

\begin{figure}[tp!]
\begin{center}
\includegraphics[scale=0.7]{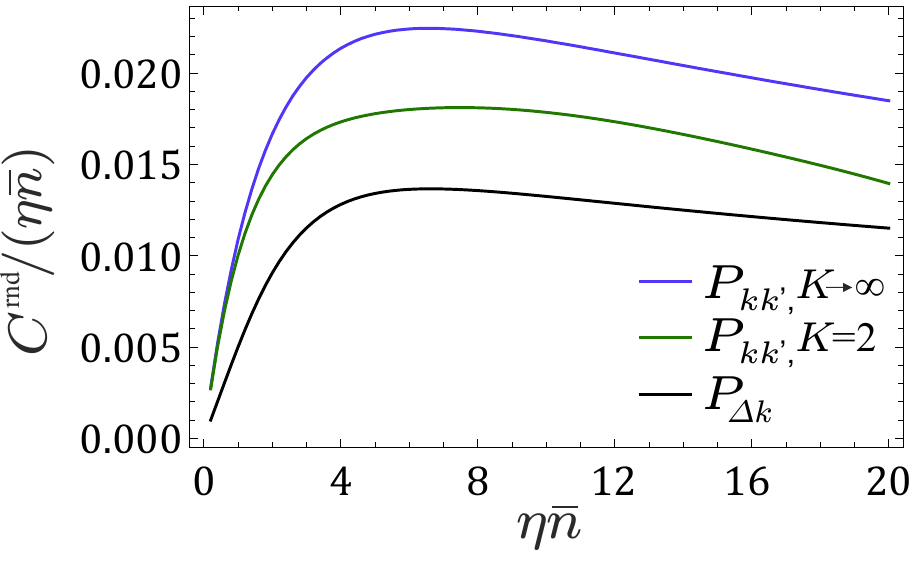}
\caption{\KB{Chernoff information per one detected photon $C^{\incoh}/(\eta\bar{n})$ as a function of the average photon number $\eta\bar{n}$  in a single realization of the visibility measurement with a random global phase, depicted for the pair of visibilities ${\cal V}_1 = 0.98$ and ${\cal V}_2 = 0.56$. The scenario based on full photocount statistics $K\rightarrow\infty$ (blue line) is compared with limited photon number resolution $K=2$ (green line), and inference based only on the photocount difference $\Delta k = k'-k$ 
(black line).}} \label{singlechernoff}
\end{center}
\end{figure}

In Fig.~\ref{singlechernoff} we depict $C^\incoh/(\eta\bar{n})$ as a function of the received optical energy $\eta\bar{n}$ for an exemplary pair of visibilities ${\cal V}_1 = 0.98$ and ${\cal V}_2 = 0.56$. The linear scaling of the ratio $C^\incoh/(\eta\bar{n})$ with $\eta\bar{n}$ for $\eta\bar{n} \ll 1$ is explained by the fact that for very weak signals detection of minimum two photons in a single realization of the measurement is necessary to obtain any meaningful information \cite{Jachura2017}. Consequently, in this regime the leading term of the Chernoff information $C^\incoh$ is proportional to $(\eta\bar{n})^2$, which gives unfavorable quadratic scaling with the channel transmission.
Beyond the two-photon regime corresponding to low optical energies, the ratio $C^\incoh/(\eta\bar{n})$ exhibits a well pronounced maximum in $\eta\bar{n}$. This observation can be used to draw the following operational conclusion. Suppose that the total optical energy available at transmitters is $\bar{n}_{\text{tot}}$. If $\bar{n}$ photons are used in a single realization of the interferometric measurement, one can afford $N= \bar{n}_{\text{tot}}/ \bar{n}$ repetitions. Let us rewrite the Chernoff bound on the error probability as $\exp(-N C^\incoh)/2 = \exp [-\eta\bar{n}_{\text{tot}}C^\incoh /(\eta\bar{n})]/2$. Assuming a fixed $\bar{n}_{\text{tot}}$, which can be taken as the overall cost of implementing the communication primitive, it is beneficial to optimize $C^\incoh/(\eta\bar{n})$ for a single realization.

Remarkably, the optimum of $C^\incoh/(\eta\bar{n})$ occurs for $\eta\bar{n}$ in the few-photon range and information needed for hypothesis testing is distributed in a non-trivial manner across the entire joint photocount statistics. To illustrate this point, in Fig.~\ref{singlechernoff} we depict also the noticeably lower ratio $C^\incoh/(\eta\bar{n})$ calculated for detection that could resolve only up to $K=2$ photocounts over the signal duration. Further, using only the marginal distribution for the photocount number difference $P_{\Delta k} ({\cal V}) = \sum_{k} P_{k,k+\Delta k}({\cal V})$ reduces significantly the Chernoff information, as also shown in Fig.~\ref{singlechernoff}. The above observations are universal as long as one of the two visibilities is sufficiently high, which is the case of quantum protocols motivating this study. In Fig.~\ref{chernoffymapa}(b) we plot the maximum $C^\incoh/(\eta\bar{n})$ as a function of the absolute values of the visibilities $|{\cal V}_1|$ and $|{\cal V}_2|$ to be discriminated between, along with the optimal average photon number that should be used in a single realization shown in Fig.~\ref{chernoffymapa}(c). Generally, the amount of Chernoff information per unit optical energy is lower than in the coherent scenario depicted in Fig.~\ref{chernoffymapa}(a), which is easily explained by the lack of the phase reference. \KB{Nevertheless}, the available information also scales linearly with the optical energy, which implies that the scaling advantage over classical protocols should be analogous to the coherent case.

\begin{figure}[t]
\begin{center}
\includegraphics[scale=0.45]{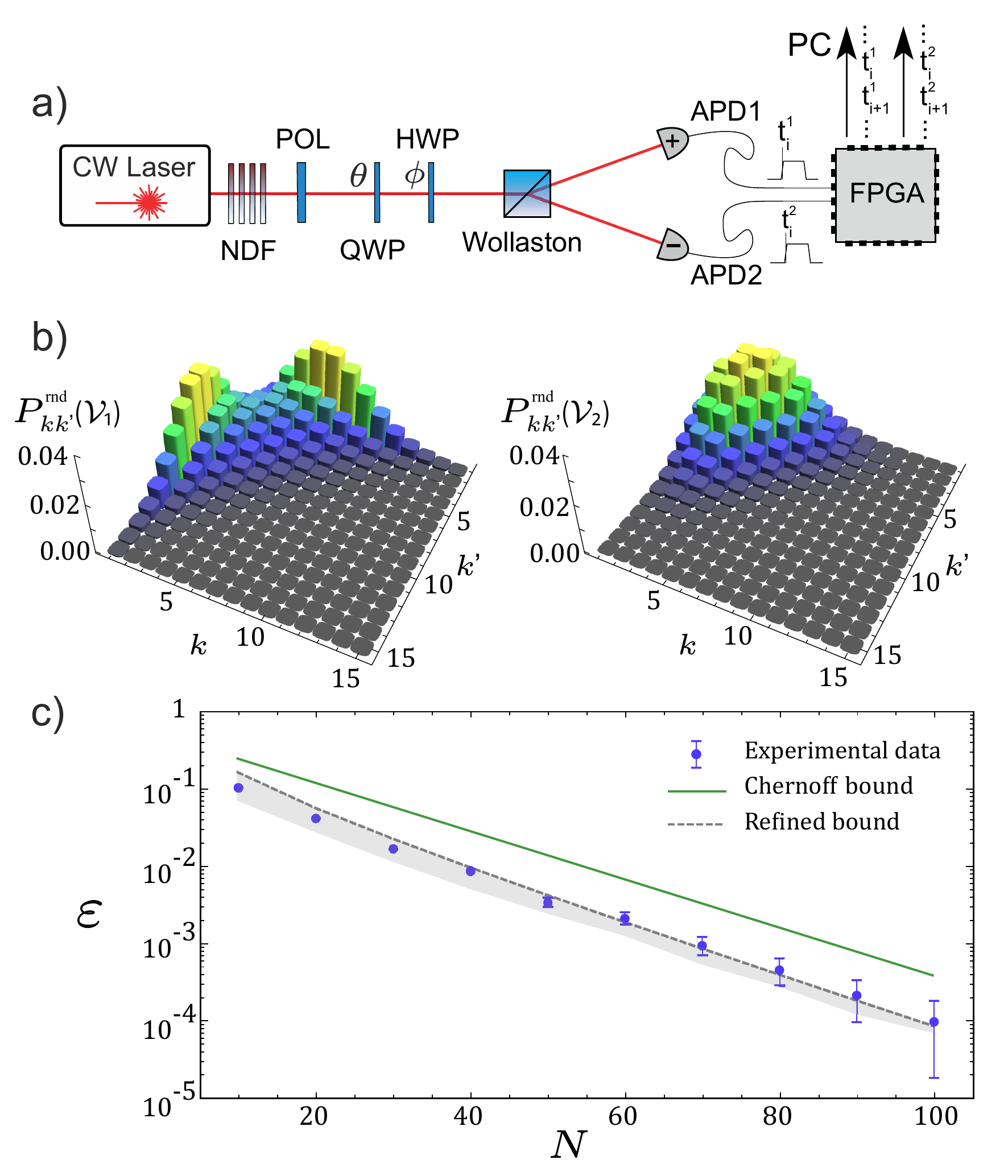}
\caption{(a) The simplified scheme of experimental setup. NDF, neutral density filters; POL, polariser; HWP, half-wave plate; QWP, quarter-wave plate; APD, avalanche photodiode; FPGA, time tagger based on field-programmable gate array. (b) Experimental joint photocount distributions $P^{\incoh}_{kk'}$ obtained from approx.\ $1.5 \times 10^{6}$ outcomes $kk'$ for visibilities ${\cal V}_1 = 0.98$ (left) ${\cal V}_2 = 0.56$ (right).
(c) The error probability $\varepsilon$ in hypothesis testing as a function of the dataset size $N$, determined for each $N$ from $1.5\times10^{4}$ repetions of the Neyman-Pearson test on independent sets of experimental data. The results are compared with the standard Chernoff bound (green, solid line) and its refined version (grey, dashed line). The errorbars account for one standard deviation. The gray shaded region corresponds to error probability of testing hypotheses $\mathcal{V}_1=0.98$ vs.\ $\mathcal{V}_2$ from the range $0\leq \mathcal{V}_2\leq 0.56$ using the Neyman-Pearson test designed for $\mathcal{V}_2=0.56$.} \label{experiment}
\end{center}
\end{figure}

We performed a proof-of-principle experimental demonstration of binary hypothesis testing for a pair of visibilities ${\cal V}_{1}=0.98$ and ${\cal V}_{2}=0.56$ using a collinear interferometric setup presented in Fig.~\ref{experiment}(a). We employed a continuous-wave $800$~nm laser diode attenuated by a series of neutral-density filters down to $\approx10^{-14}$~W of power followed by a polarizer ensuring a well-defined linear polarization. The beam is subsequently sent through a combination of a quarter- and a half-wave plate whose respective rotation angles $\theta$ and $\phi$ define the normalized intensities after the Wollaston polarizer as $I^{\pm} = (1\pm \mathrm{Re}[e^{4i\phi-2i\theta}\cos(2\theta)])/2$.
Hence our experimental setup can be viewed as a fully equivalent simulation of a standard interferometer with the complex visibility tunable in the entire phase and absolute value range by an appropriate rotation of the wave plates. To realize the random phase scenario we collected data for 50 half-wave plate angles $\phi$ uniformly probing a full period of the visibility phase. Both output beams were monitored by free-running avalanche photodiodes (APD) connected to a time tagger based on field-programmable gate array architecture, which registered photocounts with $3.3$~ns temporal resolution.

The time tagged counts for each of the two detectors were grouped over $80$~$\mu\mathrm{s}$-long time intervals. With the used input power, this interval corresponds to the mean photocount number $\eta\bar{n} = 6.3$ which gives the partitioning of the total optical energy that nearly maximizes information per one detected photon.
The numbers of photodetection events accumulated over an individual interval yield the single realization outcome $kk'$. The
$50$~ns dead time of the detectors used in the setup did not noticeably distort the measured photon statistics.
We collected approx.\ $1.5\times10^{6}$ pairs $kk'$ for each of the two visibilities. This allowed us to \KB{determine} the joint probability distributions $P^{\incoh}_{kk'}({\cal{V}}_{1,2})$ depicted in Fig.~\ref{experiment}(b), which within the resolution of the graphs match perfectly the theoretical values given by Eq.~(\ref{Eq:probability}). A detailed analysis is presented in
 Supplemental Material \cite{suplement}.

In order to experimentally determine the error probability of binary hypothesis testing one needs to repeat the \KB{test} procedure multiple times feeding it with \KB{independent sets} of experimental data obtained for \KB{a fixed visibility}. We realized this by selecting from the experimental results an ensemble of $M = 1.5\times10^{4}$ datasets [$(kk')_{1}$,...,$(kk')_{N}$]$_{1}$, [$(kk')_{1}$,...,$(kk')_{N}$]$_{2}$,...,[$(kk')_{1}$,..., $(kk')_{N}$]$_{M}$ consisting of $N$ photocount pairs. We applied the Neyman-Pearson test \cite{Cover1991} to each dataset selecting as the test result the visibility yielding a higher likelihood of photocounts group observation i.e. ${\cal V}_{1}$ if $\prod_{i=1}^{N}P^{\incoh}_{(kk')_{i}}({\cal V}_{1})>\prod_{i=1}^{N}P^{\incoh}_{(kk')_{i}}({\cal V}_{2})$ and ${\cal V}_{2}$ otherwise. The probability of error was evaluated as the ratio of erroneous hypothesis determinations to the number of groups $M$ used for testing.
That way we estimated the conditional error $\varepsilon({\cal V}_{1}|{\cal V}_{2})$
of inferring visibility ${\cal V}_1$ when ${\cal V}_2$ was the true one and the reverse error $\varepsilon({\cal V}_{2}|{\cal V}_{1})$.

In Fig.~\ref{experiment}(c) we compare the average error probability determined from experimental data $\varepsilon=[\varepsilon({{\cal{V}}_{1}}|{{\cal{V}}_{2}})+\varepsilon({{\cal{V}}_{2}}|{{\cal{V}}_{1}})]/2$ with both \KB{the standard} Chernoff bound for the random phase scenario and the refined Chernoff bound \cite{Bahadur1960} derived explicitly in Supplemental Material \cite{suplement}. In accordance with theoretical predictions, the experimental error remains below the upper bound provided by the Chernoff bound \cite{Cover1991} reaching its refined version for asymptotically large number $N$ of outcomes used for hypothesis testing \cite{Bahadur1960}.
For the fingerprinting protocol, the case of unequal inputs would hold the laxer promise of the visibility ${\cal V}_2 \le 0.56$. The shadowed grey region in Fig.~\ref{experiment}(c) indicates the range of error values obtained from Monte Carlo simulated photon count statistics with ${\cal V}_1=0.98$ and $0 \le {\cal V}_2 \le 0.56$, and processed using the Neyman-Pearson test designed for ${\cal V}_2 = 0.56$. It is seen that the decision rule works also in this more general scenario.

Let us close by discussing the parameter regime required to demonstrate quantum advantage for the fingerprinting protocol based on the primitive presented here. For input datasets $n$ bits long, in the classical scenario it is necessary to reveal at least $O(\sqrt{n})$ bits of information \cite{Ambianis1996}. As shown in Supplemental Material \cite{suplement}, in the absence of an external phase reference the strategy presented here makes it possible to maintain the exponential enhancement in the number of revealed bits scaling as $O(\log_2 n)$, analogously to the coherent protocol \cite{Arrazola2014}.
For the error probability $\varepsilon=10^{-4}$ our protocol beats the best currently known classical protocol \cite{Babai1997} for $n \ge 2.3\times 10^5$ and the ultimate classical limit \cite{Guan2016} for $n \ge 6.3\times 10^8$ bits. It is assumed here that for identical inputs the deviation of the visibility ${\cal V}_1 = 0.98$ from one stems from experimental imperfections, while unequal inputs are guaranteed to produce maximum visibility ${\cal V}_1 = 0.56$ with the same contribution from imperfections. In this scenario the attainable code rate for mapping input datasets onto binary phase patterns is $R=0.12$, which implies that the quantum advantage can be observed for pattern lengths exceeding $1.9 \times 10^{6}$ and $5.2 \times 10^{9}$ to beat the best known classical protocol and the classical limit respectively. If the optical signals are modulated with 100~GHz bandwidth available for standard $\mathrm{LiNbO}_{3}$ electro-optic modulators technology \cite{Noguchi1998}, one would require laser sources correspondingly with a kHz or a few-Hz linewidth to ensure phase stability over the signal duration. While the former requirement can be met by commercial single-frequency lasers, in the latter case more sophisticated, yet available, laser systems would be needed \cite{Stoehr2006,Notcutt2005}.


In conclusion, we described and verified experimentally a strategy to identify the modal overlap between two optical signals with a random relative phase using higher-order interference.
It can be viewed as an implementation primitive for a number of quantum-enhanced protocols, when a shared phase reference is not available. As illustrated by the quantum fingerprinting example, this approach offers analogous scaling advantage compared to classical protocols as schemes utilizing first-order coherence. The experimental demonstration of the quantum advantage should be within the reach of current technology.


We thank E. Kashefi, N. L. L\"{u}tkenhaus, F. Xu, and Q. Zhang for insightful discussions. This work was supported by the Foundation for Polish Science under the TEAM project ``Quantum Optical Communication Systems'' co-financed by the European Union under the European Regional Development Fund. M. Jachura was supported by the Foundation for Polish Science.

\newpage

\begin{widetext}
\begin{centering}
\large \textbf{Supplemental Material for "Visibility-based hypothesis testing using higher-order optical interference"}
\end{centering}
\\
\\
This document provides supplementary information to "Visibility-based hypothesis testing using higher-order optical interference". We present here derivation of the interference visibility of light in two partially overlapping modes and show how dark counts can be incorporated into the effective visibility. We also derive an analytic expression for photocounts probability in the random phase scenario and compare it with experimentally measured statistics. Additionally we derive a refined Chernoff bound for the error probability in binary hypothesis testing. Finally we describe optical quantum fingerprinting protocol for coherent and incoherent scenario and show that lack of shared phase reference does not destroy exponential advantage of communication complexity over classical protocols.
\end{widetext}

\appendix

\section{Interference visibility}

Consider two optical signals with amplitudes $\alpha$ and $\beta$, described by normalized complex waveforms in the temporal domain $u(t)$ and $v(t)$,
\begin{equation}
\int dt \, |u(t)|^2 = \int dt \, |v(t)|^2 = 1.
\end{equation}
The signals are combined at a balanced beam splitter. The time-integrated intensities at the output ports of the beam splitter  can be written as:
\begin{multline}
I_{\pm} = \frac{1}{2} \int dt \, |\alpha u(t) \pm \beta v(t)|^2 \\
= \frac{|\alpha|^2 + |\beta|^2}{2} \left[ 1 \pm \text{Re} \left( \frac{2\alpha\beta^\ast}{|\alpha|^2 + |\beta|^2} \int dt \, u(t) v^\ast(t) \right) \right]
\label{Eq:intensities}
\end{multline}
This expression has the form $I^{\pm}({\cal V})  = {\eta \bar{n}} (1 \pm \text{Re} {\cal V})/2$ given in the main text with $|\alpha|^2 + |\beta|^2 = \eta\bar{n}$ and
\begin{equation}\label{eq:visibility_uv}
{\cal V} = \frac{2\alpha\beta^\ast}{|\alpha|^2 + |\beta|^2} \int dt \, u(t) v^\ast(t).
\end{equation}
When the signals have equal amplitudes, $\alpha=\beta$, the visibility parameter ${\cal V}$ is given directly by the scalar product between the normalized signal waveforms. If the signals are misaligned at the beam splitter, e.g.\ in the transverse spatial degree of freedom, ${\cal V}$ is additionally multiplied by a spatial integral characterizing the overlap between spatial field distributions.

If the detectors employed to determine the photocount statistics exhibit dark counts characterized by Poissonian statistics with the mean $\bar{n}_{\text{dark}}$ over the signal duration, the time integrated intensities need to be replaced by
$I_{\pm} \rightarrow I_{\pm} + \bar{n}_{\text{dark}}$. It is straightforward to show that in this scenario they can also be cast into the standard form with the following substitutions:
\begin{align}
\eta\bar{n} & \rightarrow \eta\bar{n} + 2\bar{n}_{\text{dark}} \nonumber \\
{\cal V} & \rightarrow \frac{\eta\bar{n}}{\eta\bar{n} + 2\bar{n}_{\text{dark}}}{\cal V}.
\end{align}
Thus dark counts additionally reduce the effective visibility by the factor ${\eta\bar{n}}/(\eta\bar{n} + 2\bar{n}_{\text{dark}})$.

For phase-keyed signals composed of sequences of $m$ pulses with imprinted phase patterns $\phi_1^A, \phi_2^A, \ldots, \phi_m^A$ and $\phi_1^B, \phi_2^B, \ldots, \phi_m^B$, the waveforms can be written as
\begin{align}
u(t) & = \frac{1}{\sqrt{m}} \sum_{j=1}^{m} e^{i\phi_j^A} u_j(t) \\
v(t) & = \frac{1}{\sqrt{m}} \sum_{j=1}^{m} e^{i\phi_j^B} u_j(t)
\end{align}
where $u_1(t), u_2(t), \ldots, u_m(t)$ are normalized waveforms describing individual pulses in the sequence. Assuming that individual pulse waveforms are mutually orthogonal, the overlap between the signal waveforms reads
\begin{equation}
\int dt \, u(t) v^\ast(t) = \frac{1}{m} \sum_{j=1}^{m} e^{i(\phi_j^A- \phi_j^B)}.
\label{Eq:PSKOverlap}
\end{equation}

\section{Quantum fingerprinting with binary shift keyed signals}

An edifying example of a protocol to which our discrimination strategy can be directly applied is quantum fingerprinting. In the quantum fingerprinting protocol the Referee needs to decide whether $n$-bit long strings $x,y$ in possession of two separate parties Alice and Bob, are identical or different. It is known that classically the communication complexity of such a task is $O(\sqrt{n})$, i.e. both Alice and Bob need to reveal $O(\sqrt{n})$ bits to the external Referee \cite{Ambianis1996}. On the other hand it can be shown \cite{Buhrman2001} that by using quantum communication it is sufficient to reveal only $O(\log n)$ bits of information which is an exponential improvement over the classical case.

In the first step of the protocol Alice and Bob convert their strings into $m$-bit codewords using an error correcting code ${\tt E}$, which ensures that the relative Hamming distance between any two different codewords $\delta$ is greater or equal than some minimal value $\delta_{\mathrm{min}}$. The relation between lengths of the codewords $m$ and input bit strings is characterized by the rate of the code $r=n/m$. The Gilbert-Varshamov bound \cite{Lint1987} states that the maximum attainable code rate for a given $\delta_{\textrm{min}}$ is given by
\begin{equation}\label{eq:hamming_coh}
r=1-h_2(\delta_{\textrm{min}}),
\end{equation}
where $h_2(x) = - x \log_2 x - (1-x) \log_2 (1-x)$ is the binary entropy function.


In the optical realization of quantum fingerprinting, Alice and Bob send to a Referee a sequence of coherent light pulses with information about codewords encoded in the phase patterns of the light field. For instance, they may use binary phase shift keyed (BPSK) signals \cite{Arrazola2014} in which each bit of the codeword is encoded in one of two phases of coherent pulses ${\tt 0} \to 0$, ${\tt 1}\to \pi$. The Referee then interferes the sequences received from Alice and Bob on a balanced beamsplitter and monitors two output ports of the beamsplitter using single photon detectors. If the initial bit strings ${\tt x}$ and ${\tt y}$ are equal then all respective pulses of both sequences have the same phases and therefore no counts are observed in one of the output ports. On the other hand, for unequal input bit strings, some pulses have different phases which results in the possibility to observe photocounts in both output ports. Using the optical terminology, the difference between codewords affects the interference visibility. For BPSK modulation the number of pulses is equal to the length of the codewords $m$. The phase of the $j$th pulse is given by $e^{i\phi_j}=(-1)^{{\tt E}_j({\tt x})}$, where ${\tt E}_j({\tt x})$ denotes the value of the $j$th bit in the codeword ${\tt x}$, and analogously for ${\tt y}$. Plugging this in Eq.~(\ref{eq:visibility_uv}) and Eq.~(\ref{Eq:PSKOverlap}) gives the visibility
\begin{equation}\label{eq:visibility_delta}
\mathcal{V}=1-2\delta,
\end{equation}
where $\delta$ is the relative Hamming distance between $E(x)$ and $E(y)$. The fingerprinting task is thus converted into the determination of  interference visibility.

In the scenario with the random relative phase only the absolute value of visibility can be measured which leads to ambiguity $|{\cal V}| = |-{\cal V}|$ meaning that in particular identical $\delta=1$ and maximally different $\delta = 0$ codewords yield the same interference result. Ref.~\cite{Jachura2017} proposed a modification of the error correcting code by appending additional bits with the same values for any codeword. For any pair of different codewords this restricts the possible value of relative Hamming distance $\Delta$ to the range $\Delta_{\text{min}} \le \Delta \le  1-\Delta_{\text{min}}$, where $\Delta_{\text{min}}$ is expressed by the parameters of the original code as $\Delta_{\text{min}} = \delta_{\text{min}}/(1+\delta_{\text{min}})$. The cost is increased length the codewords equal to $m\to m(1+\delta_{\textrm{min}})$. Consequently, the rate of such a modified code is lower than the original code rate given in Eq.~(\ref{eq:hamming_coh}) and reads
\begin{equation}\label{eq:hamming_incoh}
R=(1-\Delta_{\textrm{min}})\left[1-h_2\left(\frac{\Delta_{\textrm{min}}}{1-\Delta_{\textrm{min}}}\right)\right].
\end{equation}
The interference visibility for modified codewords with the relative Hamming distance $\Delta$ is given by an expression analogous to Eq.~(\ref{eq:visibility_delta}):
\begin{equation}
\mathcal{V}=1-2\Delta.
\end{equation}
In the case of experimental imperfections, the right hand side should be multiplied by a factor that includes the effects of beam misalignment, dark counts, etc.

The actual task of the Referee in the fingerprinting protocol is to distinguish between two alternative hypotheses of identical inputs with associated visibility ${\cal V}_{1}$ or unequal inputs with associated visibility less or equal to ${\cal V}_{2} < {\cal V}_{1}$. Assuming that the deviation of ${\cal V}_{1}$ stems from experimental imperfections that have the same effect in the case of unequal inputs, ${\cal V}_{2}$ is given by
${\cal V}_{2} = {\cal V}_{1}(1-2\Delta_{\textrm{min}})$. As noted in the main manuscript this can be efficiently accomplished using Neyman-Pearson test designed for binary hypothesis testing taking as the second visibility the maximum allowed value $\mathcal{V}_2$. 

\begin{figure*}[tb]
\begin{center}
\includegraphics[scale=0.4]{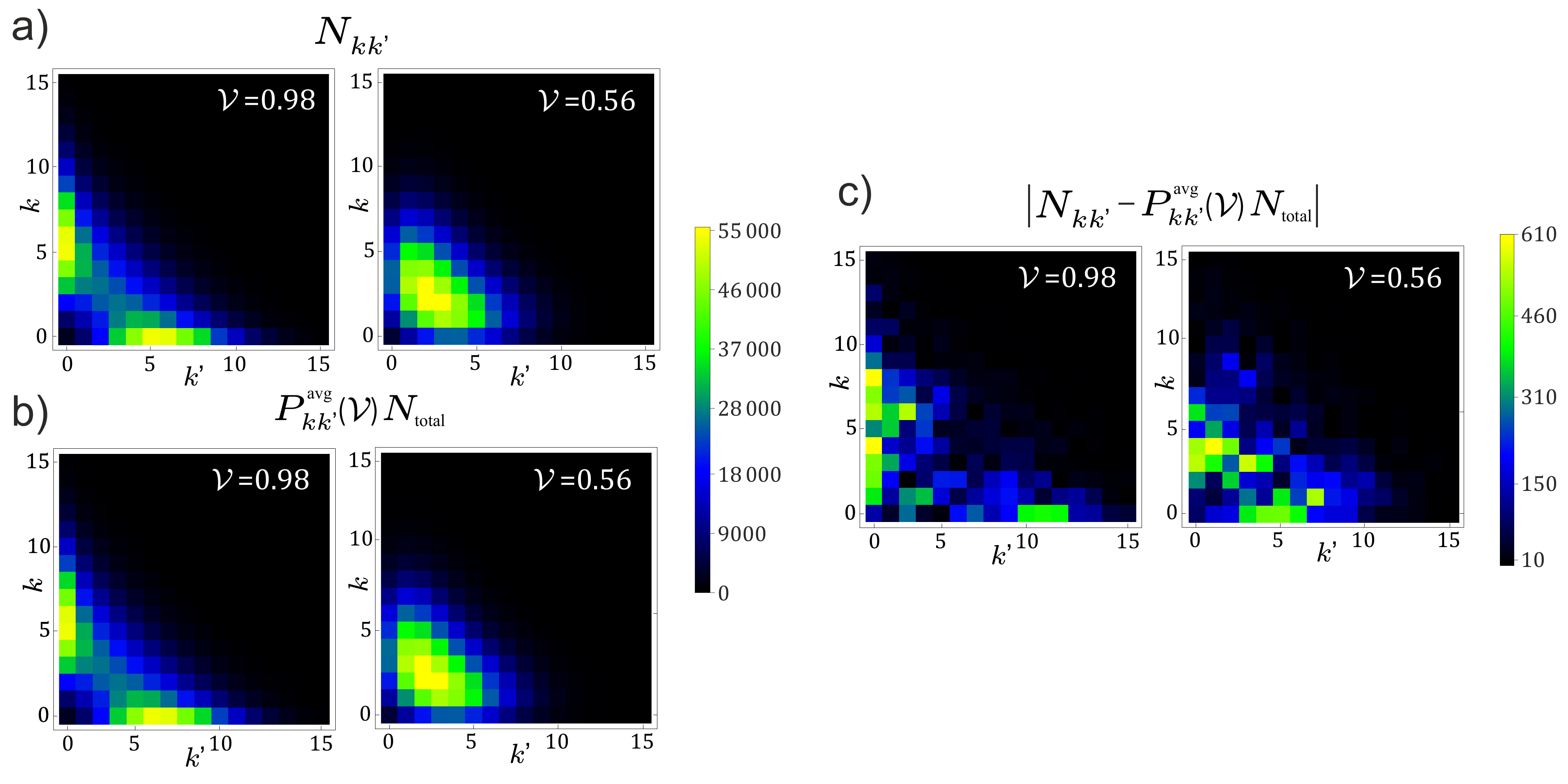}
\caption{Comparison between experimentally measured statistics of photocounts $N_{kk'}$ (a) and theoretically predicted distributions given by $P^{\text{rnd}}_{kk'}({\cal V})N_{\mathrm{total}}$ (b), where $N_{\mathrm{total}}=\sum^{15}_{k,k'=0}N_{kk'}$ and $P^{\text{rnd}}_{kk'}({\cal V})$ is defined in Eq.~\ref{Eq:finalexpressionforP}. The point-to-point difference between experimental statistics and the theoretical distribution presented in (c) does not exceed measurement uncertainty. }\label{fig:figs1}
\end{center}
\end{figure*}

\section{Experimental vs Theoretical joint photocounts distribution}

Below we shall derive the explicit expression for theoretical phase-averaged distribution of joint photocounts $P^{\text{rnd}}_{kk'}({\cal{V}})$ as well present its quantitative comparison with experimental data.

Let us begin with the formula describing the output intensities $I^{\pm}$ for the visibility ${\cal V} = e^{i\varphi}|{\cal V}|$
\begin{equation}
I^{\pm}({\cal V}) = \frac{\eta \bar{n}}{2} (1+|{\cal V}|\cos\varphi)
\label{intensitiespm}
\end{equation}
The phase-averaged distribution is given by the integral:
\begin{equation}
P^{\text{rnd}}_{kk'}({\cal V}) =  \int_{0}^{2\pi}  \frac{d\varphi}{2\pi} \, p^{+}_{k}(e^{i\varphi}{\cal V}) p^{-}_{k'}(e^{i\varphi}{\cal V}).
\end{equation}
Since the count statistics on one photodetector is given by a Poissonian distribution $p_{k}^{\pm}({\cal V}) = \exp[-I^\pm({\cal V})] {[I^{\pm}({\cal V})]^{k}}/{k!}$ the joint phase-averaged distribution can be expressed as:
\begin{multline}
P^{\text{rnd}}_{kk'}({\cal V}) = \int_{0}^{2\pi}  \frac{d\varphi}{2\pi} \, \frac{1}{k!k'!}[I^{+}({\cal V})]^{k}[I^{-}({\cal V})]^{k'}\\\times e^{-I^{+}({\cal V})-I^{-}({\cal V})}.
\label{statist}
\end{multline}
After plugging the explicit formulas for the intensities (\ref{intensitiespm}) into the integrand, the RHS of Eq.~\ref{statist} becomes:
\begin{multline}
\frac{e^{-\eta\bar{n}} (\eta\bar{n})^{k+k'}}{2^{k+k'}k!k'!}\int_{0}^{2\pi}  \frac{d\varphi}{2\pi} \, (1+|{\cal V}|\cos\varphi)^{k}\\\times (1-|{\cal V}|\cos\varphi)^{k'},
\end{multline}
which after applying binomial expansions can be written as:
\begin{multline}
\frac{e^{-\eta\bar{n}} (\eta\bar{n})^{k+k'}}{2^{k+k'}k!k'!} \int_{0}^{2\pi}  \frac{d\varphi}{2\pi} \, \sum^{k}_{m=0}\sum^{k'}_{n=0} \binom{k}{m} \binom{k'}{n} \\ \times(-1)^{n}|{\cal V}|^{m+n} (\cos\varphi)^{m+n}.
\end{multline}
We can simplify the expression by reordering the integral and the double-sum:
\begin{multline}
P^{\text{rnd}}_{kk'}({\cal V}) =  \sum^{k}_{m=0}\sum^{k'}_{n=0} \frac{e^{-\bar{n}\eta} (\eta\bar{n})^{k+k'}(-1)^{n}|{\cal V}|^{m+n}}{2^{k+k'}m!(k-m)!n!(k'-n)!}\\ \times \int_{0}^{2\pi}  \frac{d\varphi}{2\pi} (\cos\varphi)^{m+n},
\label{Eq:finalexpressionforP}
\end{multline}
The phase-averaged integer powers of cosine function can be calculated analytically yielding a compact solution:
\[ \int_{0}^{2\pi}  \frac{d\varphi}{2\pi} (\cos\varphi)^{j} = \begin{cases}
      0 &  \text{odd $j$}  \\
      \frac{1}{2^{j}}\binom{j}{j/2}& \text{even $j$}, \\

\end{cases}
\]
which makes Eq.~(\ref{Eq:finalexpressionforP}) simple enough to evaluate using symbolic computation programs such as Wolfram Mathematica.

If the joint photocount statistics is truncated up to $K$ photocounts. Since all possible pairs $kk'$ with single or both photocounts above this threshold value contribute to probabilities $P_{Kk'}$,$P_{kK}$, or $P_{KK}$, the determined distribution $P_{kk'}({\cal V})$ is given by one of four expressions:

\[ P_{kk'}({\cal V}) = \begin{cases}
      P^{\text{rnd}}_{kk'}({\cal V}) &  k,k' < K \\
      \sum^{\infty}_{k'=K}P^{\text{rnd}}_{kk'}({\cal V}) &  k < K, k' = K\\
       \sum^{\infty}_{k=K}P^{\text{rnd}}_{kk'}({\cal V}) &  k' < K, k = K \\
       \sum^{\infty}_{k,k'=K}P^{\text{rnd}}_{kk'}({\cal V}) &  k' = K, k = K,\\
       \end{cases}
\]
such that $ \sum_{k,k'=0}^{K}P_{kk'}({\cal V}) = 1$. To avoid information loss resulting from the distribution truncation we carefully adjusted the beam intensity and counts grouping time to resolve virtually all photocounts.

In Fig.~\ref{fig:figs1} we present a comparison between experimentally measured numbers $N_{kk'}$ of observed photocount pairs $kk'$ Fig.~\ref{fig:figs1}(a) and the theoretical prediction given by $P^{\text{rnd}}_{kk'}({\cal V})N_{\mathrm{total}}$ Fig.~\ref{fig:figs1}(b), where $N_{\mathrm{total}}=\sum^{15}_{k,k'=0} N_{kk'}$ stands for the total number of registered pairs. The point-to-point difference between experimental statistics and the theoretical distribution presented in Fig.~\ref{fig:figs1}(c) does not generally exceed two standard deviations of measured counts number given by $\sqrt{N_{kk'}}$.

\section{Refined Chernoff bound}

Here we will present a derivation of a refined version of the Chernoff bound following from \cite{Bahadur1960}. Assume that we want to distinguish between two equiprobable hypotheses, characterized by random distributions $p(x)$ and $q(x)$, based on $N$ repetitions of the experiment in a way that minimizes the average error. If we perform Neyman-Pearson test \cite{Cover1991} we choose as the correct one the hypothesis that yields larger likelihood. The average probability of error is then given by
\begin{multline}
\varepsilon=\frac{1}{2}\left(\text{Pr}[p(x_1)\dots p(x_N)\geq q(x_1)\dots q(x_N)|q]+\right.\\ +\left.\text{Pr}[q(x_1)\dots q(x_N)\geq p(x_1)\dots p(x_N)|p]\right),
\end{multline}
where $\text{Pr}[q(x_1)\dots q(x_N)\geq p(x_1)\dots p(x_N)|p]$ denotes the probability that we will obtain values $x_1,\dots, x_N$ for which $q(x_1)\dots q(x_N)\geq p(x_1)\dots p(x_N)$ assuming the correct hypothesis is given by $p(x)$, and analogously for $\text{Pr}[p(x_1)\dots p(x_N)\geq q(x_1)\dots q(x_N)|q]$. Let us consider the first probability, i.e. $\text{Pr}[q(x_1)\dots q(x_N)\geq p(x_1)\dots p(x_N)|p]$. We may take the condition $q(x_1)\dots q(x_N)\geq p(x_1)\dots p(x_N)$ and divide both sides by the right hand side
\begin{equation}
\frac{p(x_1)}{q(x_1)}\dots \frac{p(x_N)}{q(x_N)}\leq 1,
\end{equation}
which after taking a natural logarithm and dividing by $N$ translates into
\begin{equation}\label{eq:condition1}
\frac{1}{N}\left(\log\frac{p(x_1)}{q(x_1)}+\dots +\log\frac{p(x_N)}{q(x_N)}\right)\leq 0.
\end{equation}
Let us introduce a new random variable $y=f(x)=-\log\frac{p(x)}{q(x)}$. Since $x$ is distributed according to $p(x)$, $y$ is distributed according to an induced probability distribution $P_y(y)=p(f^{-1}(y))$. The moment generating function of $y$ is given by
\begin{multline}\label{eq:generating_function}
\varphi_y(t)=\sum_{y} e^{t y} P_y(y)=\sum_{y} e^{t y} p(f^{-1}(y))=\\=\sum_x q(x)^tp(x)^{1-t}.
\end{multline}
We may now write Eq.~(\ref{eq:condition1}) as a condition for the mean value of $y$
\begin{equation}
\frac{1}{N}\sum_{i=1}^N y_i\geq 0.
\end{equation}
According to \cite{Bahadur1960} the probability that a mean value is larger than $0$ can be written as
\begin{multline}\label{eq:prob_second_order}
\text{Pr}\left(\frac{y_1+\dots+y_N}{N}\geq 0\right)=\\=\rho^N\left(\frac{1}{\sqrt{2\pi N}\gamma}+O\left(N^{-3/2}\right)\right),
\end{multline}
where $\rho=\varphi_y(\tau)$ and $\gamma=\sigma\tau$, where $\sigma^2=\varphi_y''(\tau)/\varphi_y(\tau)$ and $\tau$ is defined implicitly by the equation $\varphi_y'(\tau)/\varphi_y(\tau)=0$. Using Eq.~(\ref{eq:generating_function}) we may rewrite the definition of $\tau$ as
\begin{multline}
\frac{\varphi_y'(\tau)}{\varphi_y(\tau)}=\sum_x p^{*}(x)\log\frac{q(x)}{p(x)}=\\
=\sum_x p^{*}(x)[\log q(x) -\log p^{*}(x)+\log p^{*}(x)-\log p(x)]=\\
=D(p^{*}||p)-D(p^{*}||q)=0,
\end{multline}
where $p^{*}(x)=p(x)^{1-\tau}q(x)^\tau/(\sum_x p(x)^{1-\tau}q(x)^\tau)$ and $D(p||q)=\sum_x p(x)\log\frac{p(x)}{q(x)}$ is the relative entropy. The solution to the above equation is given by the coefficient $\tau=\alpha^{*}$ optimizing the Chernoff information $C=\min_{0\leq\alpha\leq 1}\log \sum_{x}q(x)^\alpha p(x)^{1-\alpha}$ \cite{Cover1991}. Using this result, other quantities required in Eq.~(\ref{eq:prob_second_order}) are given by
\begin{equation}
\rho=e^{-C},\quad \sigma^2=\sum_x p^{*}(x)\left(\log\frac{p(x)}{q(x)}\right)^2
\end{equation}
Repeating calculations for the second probability, $\text{Pr}[q(x_1)\dots q(x_N)\geq p(x_1)\dots p(x_N)|p]$, and adding the results we eventually obtain a refined bound on the average error probability
\begin{equation}
\varepsilon=\frac{1}{\sqrt{2\pi N}}\frac{e^{-NC}}{2\alpha^{*}(1-\alpha^{*})\sigma}.
\end{equation}

\section{Quantum advantage}

\begin{figure}[t!]
\includegraphics[width=\columnwidth]{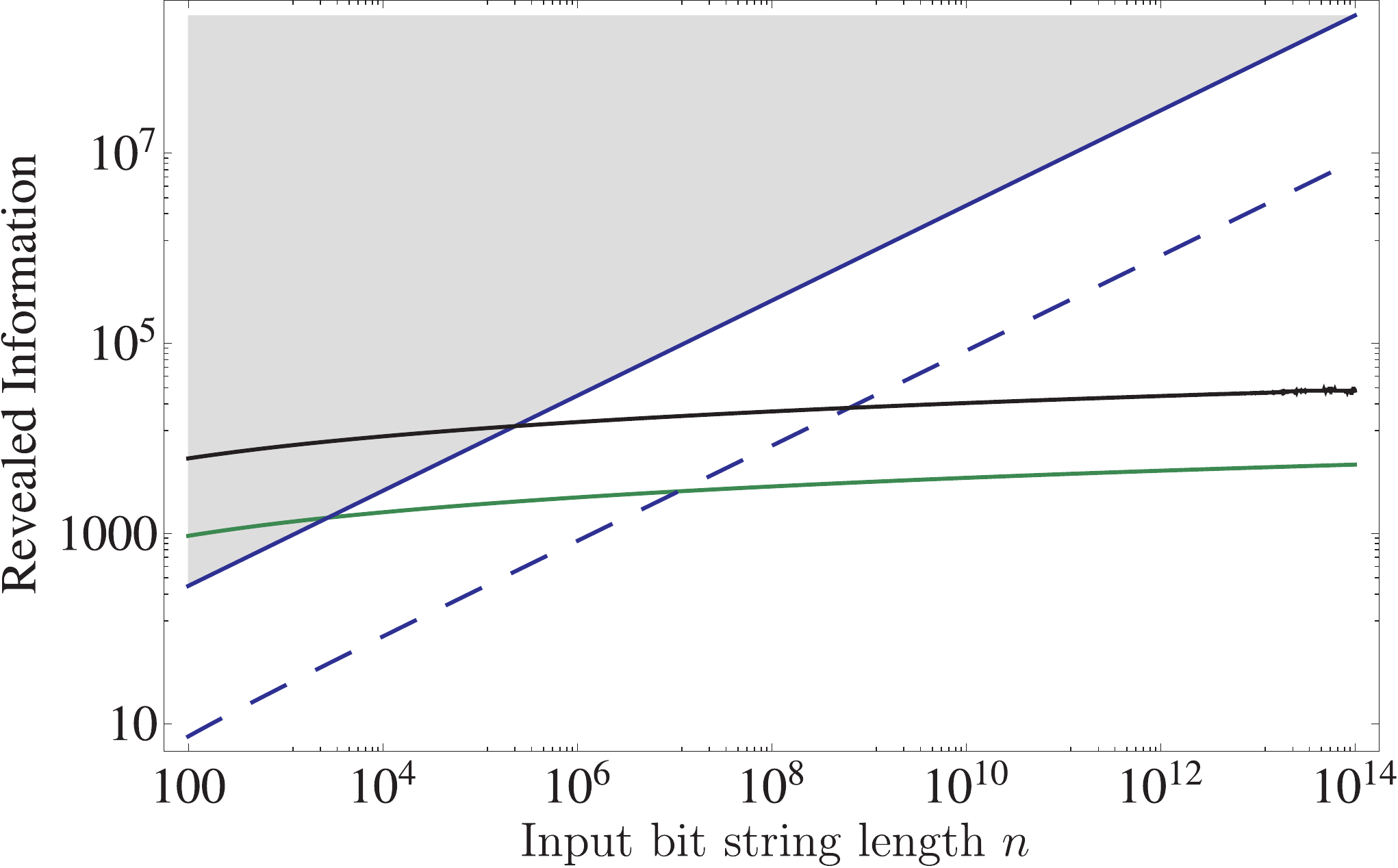}
\caption{Information revealed in a fingerprinting protocol as a function of input bit string length $n$ for the error probability $\epsilon=10^{-4}$ and visibilities $\mathcal{V}_1=0.98$ and $\mathcal{V}_2=0.56$. Coherent protocol - green; incoherent protocol - black; lower bound on any classical protocol - blue, dashed; best known classical protocol - blue, solid. Gray shaded region represents the performance worse than the best known classical protocol.}\label{fig:figs2}
\end{figure}

The amount of information revealed by Alice and Bob in coherent and incoherent quantum protocols using BPSK is upper bounded by the capacity of a lossless bosonic $2m$-mode channel with $\bar{n}$ average number of photons \cite{Giovannetti2004}, where $m$ is the number of pulses used by Alice and Bob each. Note that since code rates for both protocols are different and given by Eq.~(\ref{eq:hamming_coh}) and Eq.~(\ref{eq:hamming_incoh}), the pulse sequence length for the incoherent protocol differs from that in the coherent scenario. For large $m$ the capacity is approximately equal to $\bar{n}\log_2 m$. Since $m$ is proportional to the length of the input string $m=n/R$ for long input strings the revealed number of bits scales like $O(\log_2 n)$ also for the protocol with the random relative phase. In the following we will assume unit channel transmission, $\eta=1$. 

In Fig.~\ref{fig:figs2} we plot the revealed information in various fingerprinting protocols. The assumed probability of error is $\varepsilon=10^{-4}$ and visibilities are the same as in the main text of the article $\mathcal{V}_1=0.98$ and $\mathcal{V}_2=0.56$ which corresponds to the minimum Hamming distance $\delta_{\textrm{min}}=\Delta_{\textrm{min}} = 0.21$ for both coherent and incoherent protocols respectively. The respective code rates are given by $r=0.25$ and $R=0.12$. The average number of photons in the signal for the incoherent protocol was taken to be $\bar{n}=6.6$ which maximizes Chernoff information per photon. It is seen that both coherent and incoherent protocols have the same complexity scaling $O(\log_2 n)$, although naturally the former one requires less information to be revealed. To compare our scheme with a classical scenario, in the plot we also present two standard benchmarks i.e. a lower bound on any possible classical protocol \cite{Guan2016} and the actual performance of the best known classical protocol \cite{Babai1997}. The former reads $I_{\textrm{cl}}=(1-2\sqrt{\epsilon})(\sqrt{\frac{n}{2\ln 2}}-1)$ while the latter is equal to $I_{\textrm{best}}=4\left\lceil \frac{1}{2}\log_2\frac{1}{\epsilon}\right\rceil\sqrt{n}$ which for our probability of error yields $I_{\textrm{best}}=28\sqrt{n}$. It is seen that incoherent protocol excels the best known classical one for about $n=2.3\times 10^5$ bits which corresponds to a sequence of about $m=1.9\times 10^6$ time bins on each side. Such operating regime can be easily attained using commercially available lasers and electro-optical modulators. Beating the lower bound on any classical protocol is more demanding as it requires at least $n=6.3\times 10^8$ bits which corresponds to sequences of $m=5.2\times 10^9$ time bins. Although this regime is still within the reach of current technology it would require a modulation rates of at least tens of GHz, and significantly longer laser coherence times \cite{Stoehr2006,Notcutt2005}.

\end{document}